\documentclass{ws-procs9x6}

\newcommand{\rem}[1]{}
\newcommand{\geff}{g_{\rm eff}}
\newcommand{\vac}{{\rm vac}}
\newcommand{\muMS}{\bar\mu_{\rm MS}}

\newcommand{\0}{\over }
\def\g{g_{\rm eff}}

\begin{document}

\title{Thermodynamics of Deconfined QCD at Small and Large
Chemical Potential}

\author{Andreas IPP
\footnote{{\uppercase{W}ork supported by the \uppercase{A}ustrian \uppercase{S}cience \uppercase{F}oundation \uppercase{FWF}, project no. 16387-\uppercase{N}08.}}
}

\address{Institut f\"ur Theoretische Physik, Technische Universit\"at Wien\\
Wiedner Hauptstr. 8-10/136, A-1040 Vienna, Austria\\
E-mail: ipp@hep.itp.tuwien.ac.at}

\maketitle

\abstracts{
We present large $N_f$ QCD/QED as a test bed for improved pressure calculations, 
show how to apply the hints obtained on optimized renormalization scales at large $N_f$ to finite $N_f=2$,
and compare the results to recent lattice data.
}

\section{Introduction}

In the deconfined phase of QCD, strict perturbative calculations 
of thermodynamic potentials show poor convergence when approaching the phase transition. 
There have been a number of attempts to overcome this problem by reorganization or partial
resummation of the perturbative expansions, like HTL perturbation theory\cite{Andersen:1999va}
or $\Phi$-derivable approximations 
for 2PI skeletons\cite{Blaizot:1999ip},
but so far independent verification of these models was only possible through
lattice simulations\cite{Boyd:1996bx}.
It is therefore instructive to consider the exactly solvable special case of large number of 
flavors (large $N_f$), in which the improvements above
can be tested\cite{Moore:2002md}. 
The large $N_f$ limit can furthermore be easily 
extended to finite chemical potential\cite{Ipp:2003jy}. Also, effects relevant to full QCD, 
like the anomalous specific heat at low temperatures, 
can be readily studied in the large-$N_f$ limit\cite{Ipp:2003cj,Gerhold:2004tb}.

\begin{figure}[ht]
\centerline{\epsfxsize=4.in\epsfbox{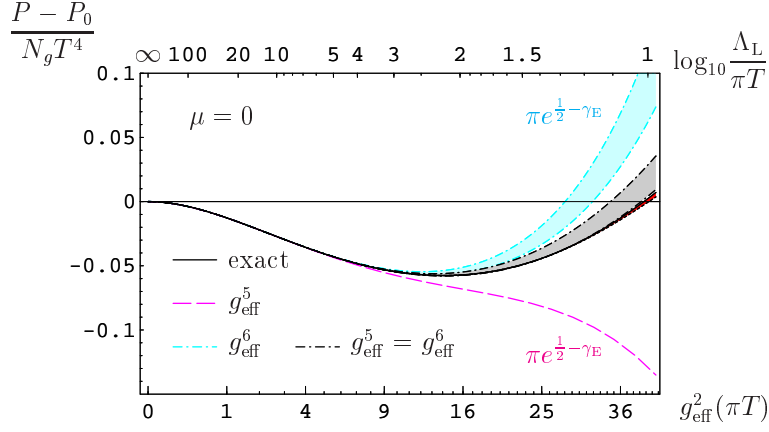}}
\caption{Exact result for the interaction pressure at zero
chemical potential 
as a function of $\g^2(\bar\mu_{\rm MS}=\pi T)$.
The 
dashed line is
the perturbative result, evaluated
with renormalization scale
$\bar\mu_{\rm MS}=\bar\mu_{\rm FAC}$;
the 
light band 
includes the numerically determined
coefficient to order $\g^6$ (with its estimated error)
also at $\bar\mu_{\rm FAC}$.
The result marked ``$\g^5=\g^6$'' 
corresponds to choosing $\bar\mu_{\rm MS}$ such
that the order-$\g^6$ coefficient vanishes and retaining
all higher-order terms contained in the plasmon term $\propto m_E^3$.
\label{fig:zeromu}}
\end{figure}

\section{Large $N_f$}

The limit of large number of flavors is formed by sending 
$N_f$ to infinity, while keeping
the combination $g^2 N_f$ as well as $N_c$ of the order of $O(1)$. 
The diagrams contributing to 
the free energy are
at leading order (LO)
only the fermion loop,
while at next-to-leading order (NLO) an infinite number of ring diagrams 
consisting of a boson loop with any number of fermion loop insertions have to be summed up\cite{Moore:2002md}.
We can treat massless QCD and ultrarelativistic QED at the same time by defining an effective coupling as 
$\geff^2 \equiv g^2 N_f / 2$ for QCD and $\geff^2 \equiv e^2 N_f$ for QED.
Large \( N_{f} \) 
contains a Landau pole of the order of \( \Lambda _{\textrm{L }}\sim \mu \exp (6\pi ^{2}/\geff ^{2}) \),
but the resulting
ambiguity for the thermal pressure at NLO is suppressed by a factor
\( ({\textrm{max}(\textrm{T},\mu )}/\Lambda _{\textrm{L}})^{4} \).

After subtracting off the vacuum part of
the ring diagrams and applying Schwinger-Dyson resummation\cite{Moore:2002md}, 
the NLO thermal pressure is given by
\begin{eqnarray}
\frac{P_{\rm NLO}}{N_g} & = & {\textstyle -{\displaystyle \int\!\frac{d^{3}q}{(2\pi)^{3}}\int_{0}^{\infty}\!\frac{dq_{0}}{\pi}}\left[2\left([n_{b}+\frac{1}{2}]\textrm{Im}\ln(q^{2}-q_{0}^{2}+\Pi_{T}+\Pi_{\vac})\right.\right.}\nonumber\\
 & & \qquad\qquad\qquad\qquad\quad{\textstyle \left.-\frac{1}{2}\textrm{Im}\ln(q^{2}-q_{0}^{2}+\Pi_{\vac})\right)}\nonumber\\
 & + & {\textstyle \left.\left([n_{b}+\frac{1}{2}]\textrm{Im}\ln(\frac{q^{2}-q_{0}^{2}+\Pi_{L}+\Pi_{\vac}}{q^{2}-q_{0}^{2}})-\frac{1}{2}\textrm{Im}\ln(\frac{q^{2}-q_{0}^{2}+\Pi_{\vac}}{q^{2}-q_{0}^{2}})\right)\right]}\,\,\,\,\end{eqnarray}
with the bosonic distribution function \( n_{b}(\omega )=1/(e^{\omega /T}-1) \) and the gauge-boson self energy functions $\Pi_T$ and $\Pi_L$. These cannot
be given in closed form except for their imaginary parts\cite{Ipp:2003qt},
but are represented by one-dimensional integrals.
We therefore have to evaluate the integrals
numerically. Parts proportional to $n_b$ can be safely integrated in Minkowski 
space,
but terms without
\( n_{b} \) are potentially logarithmically divergent. We compute them 
 by introducing a Euclidean invariant cutoff \!\cite{Moore:2002md}. 

\section{Numerical results}


Figure \ref{fig:zeromu} shows the numerical result for $\mu=0$ as a function
of $\g^2(\bar\mu_{\rm MS}=\pi T)$. 
For small coupling 
the coefficients to order $\g^6$ which are not yet known
analytically can be extracted numerically\cite{Ipp:2003jy}.
The large renormalization scale dependences of successive
perturbative approximations to order $\g^5$ beyond $\g^2\sim 4$
can be fixed by applying ``fastest apparent convergence'' (FAC)
in the $m_E^2$ parameter of dimensional reduction. Using 
$\bar\mu_{\rm MS}=\bar\mu_{\rm FAC}\equiv\pi e^{1/2-\gamma}T$
we obtain good agreement up to $\g^2\sim 9$.
The result can be further improved 
by the procedures explained below Fig.~\ref{fig:zeromu}.

\begin{figure}[ht]
\centerline{\epsfxsize=3.4in\epsfbox{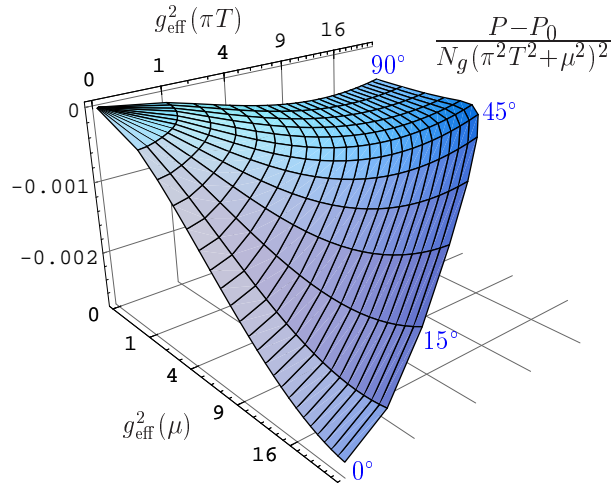}}
\caption{Exact result for the large-$N_f$ interaction pressure $P-P_0$ normalized
to $N_g(\pi^2T^2+\mu^2)^2$ as a function of $\g^2(\bar\mu_{\rm MS})$
with $\bar\mu_{\rm MS}^2=\pi^2T^2+\mu^2$
and $\phi=\arctan{\pi T\0\mu}$. \label{fig:3d}}
\end{figure}

For non-vanishing chemical potential \( \mu  \) we use the fermionic
distribution function \begin{equation}
\label{fermionicdistributionfunction}
n_{f}(k,T,\mu )=\frac{1}{2}\left( \frac{1}{e^{(k-\mu )/T}+1}+\frac{1}{e^{(k+\mu )/T}+1}\right) 
\end{equation}
which enters via the gauge boson self-energy expressions \( \Pi _{T} \)
and \( \Pi _{L} \). In Fig.~\ref{fig:3d} we display our exact results
for the interaction pressure $P-P_0 \propto N_f^0$
for the entire $\mu$-$T$ plane
(but reasonably below the scale Landau pole).
The figure shows a kink at $\phi=45^\circ$ corresponding to $\mu=\pi T$ indicating that a simple scaling behavior for the pressure at small chemical potentials would break down at larger chemical potentials\cite{Ipp:2003jy}.

\begin{figure}[ht]
\centerline{\epsfxsize=2.25in\epsfbox{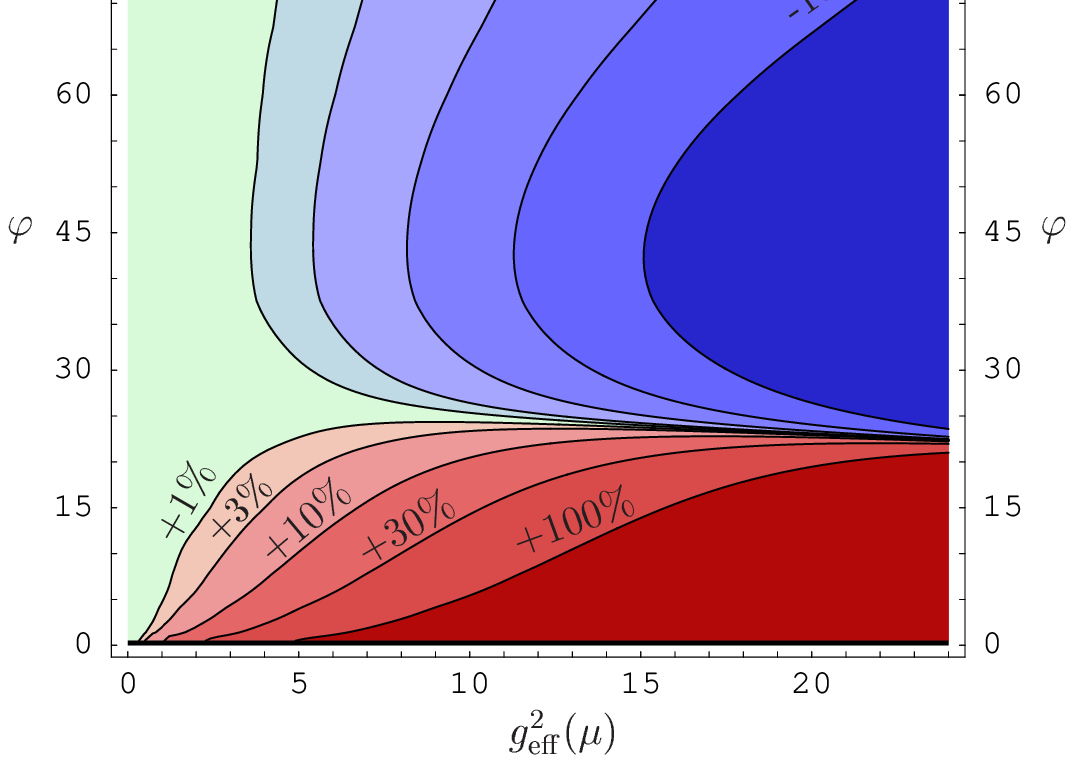}\epsfxsize=2.25in\epsfbox{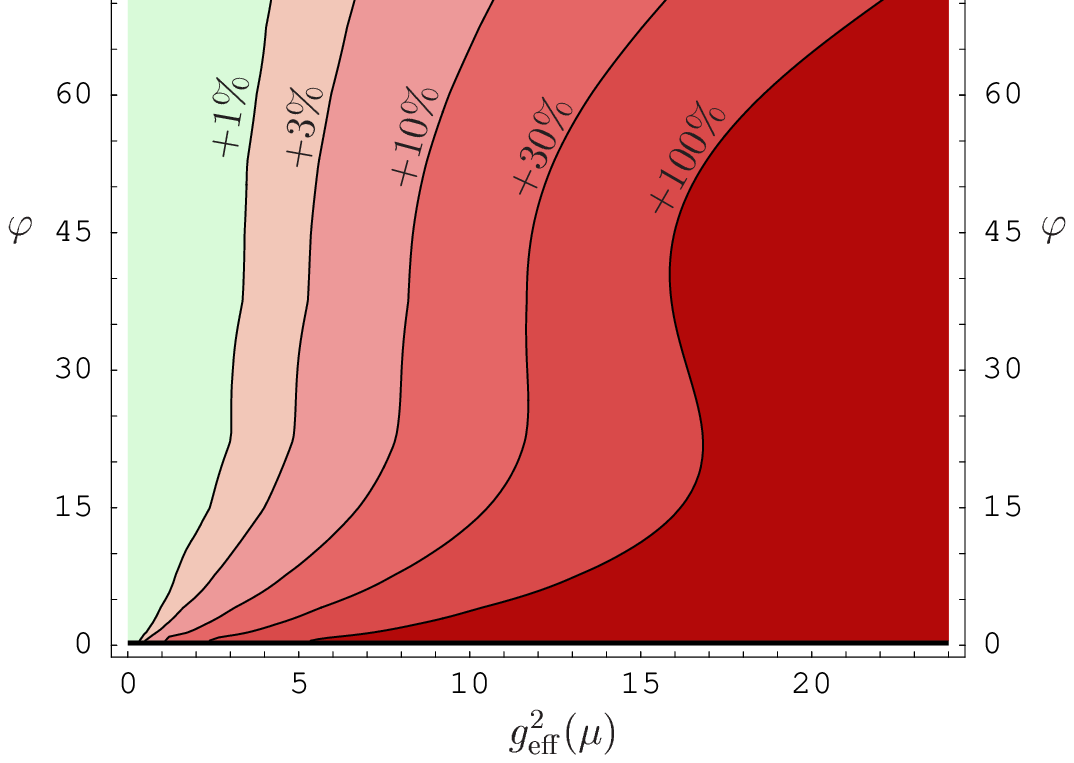}}
\caption{Percentage errors of the perturbative result for the
interaction part of the pressure to
order $\geff^5$ in the large-$N_f$ limit
as a function of $\varphi=\arctan(\pi T/\mu)$ and $\geff^2(\muMS)$ at $\muMS=\sqrt{\pi^2 T^2+\mu^2}$
for two choices of $\muMS$: Fastest apparent convergence
of $P$ as well as
$m_E^2$ (FAC-m, left panel), and of $g_E^2$ (FAC-g, right panel).
The brightest area corresponds to an error of less than 1\%,
the darkest ones to an error of over 100\%.
\label{fig:contpl}}
\end{figure}

Figure~\ref{fig:contpl} shows a comparison between the perturbative result obtained by 
dimensional reduction\cite{Vuorinen:2003fs} through order \( g^{5} \) with complete
analytic dependence on arbitrary \( T \) and \( \mu  \) to the exact result in the large $N_f$ limit.
In the left panel 
the renormalization scale is $\muMS=\muMS^{\rm FAC-m}$.
An alternative choice for the renormalization scale is to set 
$\muMS$ such that the $\geff^4$ correction
in the dimensional reduction coupling parameter $g_E^2$ 
is put to zero (FAC-g, right panel).
The accuracy of the results is comparable in both cases and decreases slowly with increasing chemical potential,
apart from an accidental zero of the error;
for $\varphi\lesssim 18^\circ$, i.e.\
$T\lesssim 0.1\mu$, the errors eventually start to grow rapidly,
marking the breakdown of dimensional reduction. 
This is precisely the region where non-Fermi-liquid effects lead to 
anomalous $T \ln T^{-1}$ terms in the entropy and specific heat\cite{Ipp:2003cj,Gerhold:2004tb}.

\begin{figure}[ht]
\centerline{\epsfxsize=3.5in\epsfbox{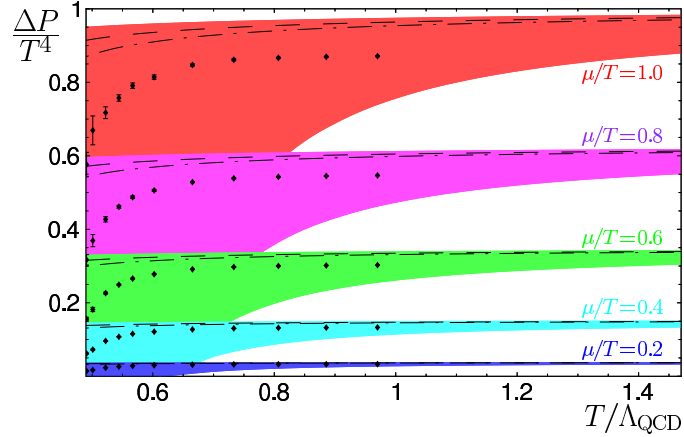}}
\caption{The difference $\Delta P=P(T,\mu)-P(T,0)$ divided by $T^4$ for $N_f=2$ using
the unexpanded three-loop result from dimensional reduction
for $\mu/T=0.2,\ldots,1.0$
(bottom to top).
Shaded areas correspond to a variation
of $\muMS$ around the FAC-m choice by a factor of 2; dashed and
dash-dotted lines
correspond to the FAC-g and FAC-m results, respectively. 
Also included are recent lattice data 
(not yet continuum-extrapolated!) assuming $T_c^{\mu=0}=
0.49\, \Lambda_{\rm QCD}$. 
\label{fig:dpr}}
\end{figure}

In Fig.~\ref{fig:dpr} we display the FAC optimized results\cite{Ipp:2003yz} 
at finite $N_f=2$ for the
difference $\Delta P=P(T,\mu)-P(T,0)$ for various
$\mu/T$ corresponding to recent lattice results\cite{Allton:2003vx}
assuming $T_0\equiv T_c^{\mu=0}=
0.49\, \Lambda_{\rm QCD}$ \cite{Gupta:2000hr}.
At $T/T_0=2$ our FAC-g and FAC-m results exceed the not-yet-continuum-extrapolated 
lattice data consistently by 10.5\% and 9\%, respectively, which is roughly the expected discretization error\cite{Karsch:2000ps}.



%
%
%
%


\begin{thebibliography}{0}

\bibitem{Andersen:1999va}
J.~O. Andersen, E.~Braaten, and M.~Strickland,  
Phys. Rev. Lett. {\bf 83} (1999) 2139--2142;
R.~Baier and K.~Redlich, Phys. Rev. Lett. {\bf 84} (2000) 2100.

\bibitem{Blaizot:1999ip}
J.~P. Blaizot, E.~Iancu, and A.~Rebhan, Phys. Rev. Lett. {\bf 83} (1999) 2906--2909;
Phys. Rev. {\bf D63} (2001) 065003;
A.~Peshier, Phys. Rev. {\bf D63} (2001) 105004.

\bibitem{Boyd:1996bx}
G.~Boyd {\em et.~al.},
Nucl. Phys. {\bf B469} (1996) 419--444;
{CP-PACS} Collaboration, M.~Okamoto {\em et.~al.}, Phys. Rev. {\bf D60} (1999) 094510.


\bibitem{Moore:2002md}
G.~D. Moore, JHEP {\bf 0210},  055  (2002);
A. Ipp, G.~D. Moore, and A. Rebhan, JHEP {\bf 0301},  037  (2003).

\bibitem{Ipp:2003jy}
A. Ipp and A. Rebhan, JHEP {\bf 0306},  032  (2003).

\bibitem{Ipp:2003cj}
A. Ipp, A. Gerhold, and A. Rebhan, 
Phys. Rev. {\bf D69},  011901  (2004).

\bibitem{Gerhold:2004tb}
A.~Gerhold, A.~Ipp, and A.~Rebhan,
hep-ph/0406087;
see also contribution in these proceedings.

\bibitem{Ipp:2003qt}
A.~Ipp,
PhD thesis TU Vienna 2003,
hep-ph/0405123.


\bibitem{Vuorinen:2003fs}
A. Vuorinen, Phys. Rev. {\bf D68},  054017  (2003).

\bibitem{Ipp:2003yz}
A.~Ipp, A.~Rebhan and A.~Vuorinen,
Phys.\ Rev.\ D {\bf 69}, 077901 (2004)

\bibitem{Allton:2003vx}
C.~R. Allton {\it et~al.}, Phys. Rev. {\bf D68},  014507  (2003).

\bibitem{Gupta:2000hr}
S. Gupta, Phys. Rev. {\bf D64},  034507  (2001).

\bibitem{Karsch:2000ps}
F. Karsch, E. Laermann, and A. Peikert, Phys. Lett. {\bf B478},  447  (2000).



\end{thebibliography}
\end{document}